\documentclass[aip,reprint]{revtex4-1}

\usepackage{color}
\usepackage[english]{babel}
\usepackage[utf8]{inputenc}
\usepackage{anysize} 
\usepackage{graphicx} 
\usepackage{amssymb} 
\usepackage{float} 
\usepackage{amsmath, amsthm, amsfonts}
\usepackage[normalem]{ulem} 

\begin{document}


\title{Hydrodynamic provinces and oceanic connectivity from a transport network 
help designing marine reserves} 



\author{Vincent Rossi,
Enrico Ser-Giacomi, 
Crist\'obal L\'opez,
and Emilio Hern\'andez-Garc\'ia,}
\affiliation{IFISC, Instituto de F\'isica Interdisciplinar y Sistemas Complejos (CSIC-UIB). Campus Universitat
Illes Balears, Crta. Valldemossa km 7.5, E-07122 Palma de Mallorca, Spain.}


\date{\today}

\begin{abstract}
Oceanic dispersal and connectivity have been identified as crucial factors for structuring marine populations 
and designing Marine Protected Areas (MPAs). Focusing on larval dispersal by ocean currents, we propose an approach 
coupling Lagrangian transport and new tools from Network Theory to characterize marine connectivity in the Mediterranean 
basin. Larvae of different pelagic durations and seasons are modeled as passive tracers advected in a simulated oceanic 
surface flow from which a network of connected areas is constructed. Hydrodynamical provinces extracted from this network 
are delimited by frontiers which match multi-scale oceanographic features. By examining the repeated occurrence 
of such boundaries, we identify the spatial scales and geographic structures that would control 
larval dispersal across the entire seascape. Based on these hydrodynamical units, we study novel connectivity 
metrics for existing reserves. Our results are discussed in the context of ocean biogeography 
and MPAs design, having ecological and managerial implications. 
\end{abstract}


\maketitle 

\section{Introduction}

Oceanic ecosystems are impacted by multiple human-induced stressors, including habitat destruction, pollution, 
overfishing and global climate change. Marine protected areas (MPAs), used for the management and conservation 
of marine ecosystems, are considered effective to mitigate some of these impacts \cite{lester2009}. 
Successful MPA design is however complicated primarily due to the difficulties in quantifying the movements 
of organisms, especially at larval stage \cite{shanks2009}, in resolving the multi-scale variability of ocean 
currents \cite{siegel2008} and in apprehending the spatial scales and biogeography of the seascape \cite{hamilton2010}. 

Marine population connectivity, i.e. the exchange of individuals among geographically separated subpopulations, 
depends on numerous factors including spawning outputs, larval dispersal, habitat availability, trophic interactions 
and adult movements \cite{cowen2009,game2009}. Among them, larval dispersal has been identified as a crucial 
factor for structuring oceanic populations \cite{cowen2006} and for determining broad-scale ecological connectivity 
\cite{treml2012}. It also plays a major role in assuring population persistence in a MPA network \cite{moffit2011}, 
especially when target species show long-distance dispersal \cite{shanks2009}. As such, patterns and magnitude of larval 
connectivity have been used to design MPAs \cite{lester2009} and assess their efficiency \cite{pelc2010}. This paper 
focuses on the dispersion of larvae by ocean currents at basin-scale, assuming they are passively transported by the 
flow (i.e. neglecting larval behavior), to inform the design of marine reserves.  

Many biophysical modeling studies \cite{cowen2006,siegel2008}, including Lagrangian approaches, 
examined marine connectivity from the so-called ``connectivity matrix'' which represents 
the probability of larval exchange between distant sites. Previous analyses were mainly limited to 
the strengths of pair-wise connections, i.e. the links from one coastal site, or MPA, to another distant one 
\cite{corell2012,vaz2013}. Another perspective to investigate connectivity is the analysis of 
dispersal network topologies \cite{treml2012,kool2013}. Recent studies applied tools derived from Graph 
Theory to document regional connectivity of near-shore MPAs in the Baltic Sea \cite{nilsson2012}, the 
Mediterranean Sea \cite{andrello2013} and in the Great Barrier Reef region \cite{thomas2014}. 
While our understanding of connectivity at small- and regional-scales has improved, previous 
efforts focused mainly on coastal/insular areas and did not provide a characterization of the seascape 
connectivity.

The significance of this shortcoming is emphasized by the growing interests for the implementation of MPAs 
in the pelagic ocean \cite{pala2013,guidetti2013} which also shelters biodiversity and important ecological 
processes \cite{game2009,kaplan2010a}. Designing open-ocean MPAs is challenging partly because larval 
connectivity and pelagic habitats are difficult to assess in such vast and dynamic environment. 

Here we use an approach coupling Lagrangian modeling and new tools from Network Theory \cite{newman2010} 
to characterize marine connectivity at basin-scale in the Mediterranean Sea. Larvae of different Pelagic 
Larval Durations (PLD) are modeled as passive Lagrangian particles advected in a simulated oceanic surface flow 
from which a network of connected areas can be constructed. Hydrodynamical provinces extracted from this transport 
network are delimited by frontiers which match mesoscale and regional oceanographic features. We then identify the 
spatial scales and structures of larval dispersal across the entire seascape and analyse connectivity metrics 
for the existing Mediterranean MPAs. We finally discussed the usefulness of our results for the design of 
marine reserves and the characterization of oceanic biomes.

\section{Materials and Methods}

\subsection{Oceanic transport and connectivity from a Network Theory approach}

The basic ingredients are (i) the tracking of passive Lagrangian particles (a model for larval transport) 
and (ii) the construction and analysis of a network of flow-mass transport. We study dispersal 
processes in the ocean based on a transport network in which:  
\begin{itemize}
 \item a \textit{node} corresponds to a geographical sub-area of the oceanic surface, 
 \item a \textit{link}, or \textit{edge}, symbolizes an effective mass transport driven by ocean currents 
 between 2 sub-areas during a given time interval. 
\end{itemize}
The transport network is thus composed of an ensemble of nodes (sub-areas), covering the entire oceanic domain 
of interest, which are inter-connected by a number of links (transport pathways). Each link is \textit{directed} 
in accord with the effective direction of the flow, and \textit{weighted} proportionally to the amount of water 
flowing from one node to another. Many tools of Network Theory were designed to examine both local 
and global properties of such network \cite{newman2010}, allowing us to explore geophysical flows 
and connectivity in a new fashion.

The Mediterranean Sea, a quasi-closed basin with its own physical circulation and ecological functioning under 
important human pressure \cite{mermex2011}, constitutes a natural laboratory for this study. In addition, it 
shelters already a hundred MPAs (whose locations were downloaded from the MedPan database) implemented for protection 
and conservation purposes. In this context, we aim at partitioning the surface Mediterranean seascape in 
hydrodynamical provinces, i.e. a set of oceanic subregions in which larvae/particles are much more likely to disperse 
efficiently within each other than among them at a given time-scale. This spatial subdivision in provinces is tantamount 
to detecting \textit{communities} within the hydrodynamical network \cite{newman2010}.

\subsection{Lagrangian bio-physical modelling}

The Lagrangian approach is a natural perspective to characterize transport phenomena affecting free-swimming 
larvae \cite{corell2012,vaz2013}. Particles are advected in an eddy-resolving velocity field generated by
the Nucleus for European Modeling of the Ocean hydrodynamical model implemented in the Mediterranean at 
a $1/16$ $deg$ horizontal resolution \cite{oddo2009}, (see also SI-text01). We focus on the upper-ocean dynamics 
over years $2002-2011$ with the use of daily horizontal flow fields at $8$ m depth (Fig. SI-1), representing the 
surface mixed layer in which larvae are assumed to be homogeneously distributed. 

Horizontal trajectories are simulated by integrating the velocity field, bi-linearly interpolated at 
any sea point, using a \textit{Runge-Kunta 4} algorithm with a time step of $1$ day, matching the resolution of 
the simulated currents. Lagrangian particles are dispersed 
as two-dimensional passive drifters \cite{corell2012,andrello2013}. Note that due to the non-fully incompressible horizontal flow field, 
vertical velocities may become significant in regions of strong divergence (e.g. coastal upwelling) and convergence (e.g. 
deep water formation). Neglecting vertical movements is however a reasonable assumption here because most particles remain 
in the selected layer over short time-scales ($\le 2$ months) since horizontal velocities are several orders of magnitude 
higher than vertical ones \cite{dovidio2004}. Another simplification is the passive character of the particles, the 
implementation of more complex larval behavior (e.g. vertical migration, mortality, settlement) being envisaged for 
future work. Under these assumptions, larval dispersal is modulated by the PLD, the period of spawning and 
the time-varying oceanic circulation.

Initial ($t_{0}$) and integration ($\tau \sim$ PLD) times are chosen according to the typical 
biological traits of marine organisms. Given the limited knowledge of their life cycles \cite{shanks2009}, 
we investigate basin-scale larval connectivity from an ecosystem-based approach \cite{coll2012,guidetti2013},
rather than focusing on a particular target species. To do so, we retain two different PLDs ($\tau = 30, 60$ days) 
and consider winter ($t_{0}$ = $1^{st}$ Jan.) and summer ($t_{0}$ = $1^{st}$ Jul.) spawning 
\cite{macpherson2006,shanks2009,andrello2013} over the years $2002-2011$. These modeling choices are 
ecologically meaningful for a number of Mediterranean organisms, especially those with wide geographical 
range and potential for large-distance dispersal (SI-text02). Sensitivity of our results 
to the parameter $\tau$ was tested by performing additional simulations for $\tau = 45$ days. 
A total of 60 factorial experiments (with starting times covering $2$ seasons over $10$ years, 
for each of the $3$ PLDs) allow the construction of $60$ connectivity matrices from which hydrodynamical provinces 
are extracted (sect. \ref{snapshotprovinces}). They are then temporally-averaged to describe robust spatial 
patterns of larval connectivity in the entire Mediterranean basin (sect. \ref{climatoprovinces}), 
finally interpreted in the context of MPA design (sect. \ref{mpaconnectivity}). 

\subsection{Construction and analysis of a transport network}

\subsubsection{Connectivity matrix}

The nodes of the transport network are defined by discretizing the surface ocean into $3270$ quasi-square boxes 
of $1/4^{o}$ horizontal-resolution (Fig. SI-2), allowing the consideration of 
important mesoscale features of the Mediterranean circulation \cite{mermex2011}. This 
procedure and the numerical diffusion it introduces \cite{froyland2003} are detailed 
in SI-text01.

$500$ Lagrangian particles evenly distributed in each oceanic box are advected, corresponding 
to a total of $\sim$ 1.6 millions trajectories in the Mediterranean basin for each experiment. The 
connectivity matrix is built using the initial and final positions of these particles. 
Each matricial element $\mathbf{P}^{t_{0},\tau}_{ij}$, i.e. the link between nodes $i$ and $j$, 
is proportional to the fraction of particles leaving box $i$ at time $t_{0}$ and arriving in box $j$ at 
time $t_{0} + \tau$. $\mathbf{P}^{t_{0},\tau}_{ij} = {\#\,particles\,from\,box\,i\,to\,box\,j}/{N_{i}} \in [0,1]$
is interpreted as the probability for a particle selected randomly in box $i$ at the initial time ending up in box $j$ at 
the final time. $N_{i}$ is a local normalization coefficient equal to the number of particles still within the 
oceanic domain after integration. Due to numerical limitations, some trajectories may indeed abort prematurely 
with the ``beaching'' of particles onto land areas. With this normalization, which concerns less than $5 \%$ of all 
particles for $\tau = 60$ days, the stochasticity of the matrix (i.e. its rows sum to 1) is ensured 
and the constraint of mass conservation fulfilled.

\subsubsection{Community detection}

To locate coherent regions in time-varying flows, the transport matrix approach has been proposed
\cite{froyland2003}. Such objects are extracted by examining eigenvalues \cite{nilsson2012} or singular 
vectors \cite{froyland2012} of the transport (connectivity) matrix which approximates the continuous advection 
operator. These concepts, along with other community detection algorithms, were recently used to 
study ecological connectivity \cite{nilsson2012,thomas2014}. 

Here we analyze the topology of the transport network to subdivide the surface ocean in hydrodynamical provinces. 
Based on the connectivity matrix $\mathbf{P}^{t_{0},\tau}$, an equivalent of the network adjacency matrix, we detect 
communities using the \textit{Infomap} algorithm \cite{rosvall2008}. 
Random walkers are considered to move in the network according to the statistical description of the surface flow 
contained in the connectivity matrix. From these synthetic trajectories, and using information theory concepts, 
\textit{Infomap} decomposes the network into a number of communities that define oceanic provinces well 
connected internally but with minimal exchanges of particle between them. The method 
is described and compared with other alternative in SI-text03. 

To evaluate the significance of the spatial partitioning, we define a coherence ratio \cite{froyland2012} 
associated with each province $k$ by: 
$\rho_{k} = {\sum_{i,j \in \mathcal{I}_{k}} N_{i} \mathbf{P}^{t_{0},\tau}_{ij}}/ \sum_{i \in \mathcal{I}_{k}} N_{i} $ 
where $\mathcal{I}_{k}$ is the set of indices that identify all boxes constituting
province $k$. Physically, $\rho_{k} \in [0,1]$ is interpreted as the fraction of particles initially released 
within a given province which remained in it after integration. Its complement $1-\rho_{k}$ measures 
the proportion of particle leaking across the boundaries of each province.

\section{Results and discussion}

\subsection{Time-dependent hydrodynamical provinces}
\label{snapshotprovinces}

\begin{figure}
\noindent
\begin{center}
\includegraphics[width=0.5\textwidth, clip=true]{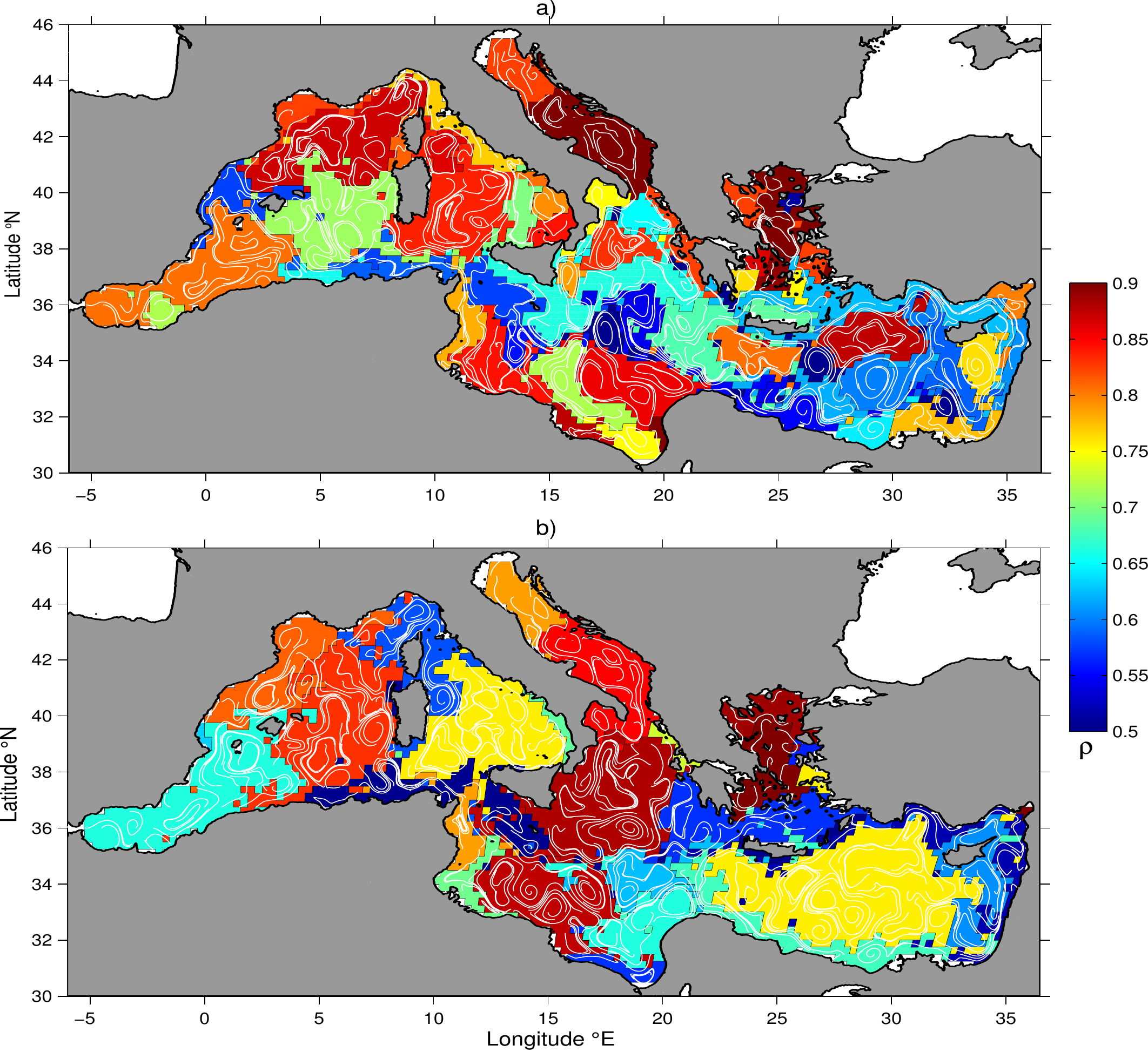}
\caption{Hydrodynamical provinces extracted from the connectivity matrices of a) Winter 2011 ($t_{0} = 1^{st}$ Jan.) 
using $\tau \simeq PLD = 30$ days and b) Summer 2011 ($t_{0} = 1^{st}$ Jul.) using $\tau \simeq PLD = 60$ days. 
Each province is colored according to its $\rho$ value (ranging from 0.5 to 0.9). White streamlines represent 
the simulated flow averaged over the period of integration (i.e. a) $1^{st}$-$30^{th}$Jan. 2011 and 
b) $1^{st}$Jul.-$29^{th}$Aug. 2011).}
\label{fig:snapshots}
\end{center}
\end{figure}

The provinces and their boundaries are dynamical objects that evolve in space and time with different 
dimensions, shapes and locations (e.g. Fig. \ref{fig:snapshots}), due to the important variability 
of the ocean circulation \cite{millot2005}. The method captures an 
elevated number of communities in the network, with 65 provinces using a PLD $=30$ days and only 32 for 
PLD $=60$ days on the exemplary calculations displayed on Fig. \ref{fig:snapshots}. Intuitively, 
the longer the tracking time, the lower the number of provinces detected and the larger 
their mean area. On average over the ensemble of experiments, community detection results in 61, 46, 36 
provinces characterized by a mean area of $4.12 \times 10^{4}$, $5.5 \times 10^{4}$, $6.8 
\times 10^{4}$ (in $km^{2}$) for $\tau = 30,\,45,\,60$ days, respectively. Because of the time-varying 
flow \cite{siegel2008}, both release time and tracking duration (simulating respectively the initiation and 
duration of the pelagic larval phase) affect the spatial partitioning. 

Most province boundaries match very well the mean flow streamlines (Fig. \ref{fig:snapshots}), suggesting 
high oceanographic relevance. While isolated streamlines are found in the cores of provinces, dense ones usually 
coincide with the detected boundaries. Hydrodynamical provinces are delimited by intense oceanic mesoscale 
structures such as jets, meanders, fronts and eddies. These features, which influence the topology of 
the transport network and thus the community detection, were recently reported to strongly impact connectivity 
\cite{vaz2013}. For instance, some mesoscale eddies are extracted as quasi-circular single provinces (e.g. in the 
Alboran Sea and in the southern Levantine basin), in good agreement with the flow streamlines (Fig. \ref{fig:snapshots}a). 
Other mesoscale structures are contained in larger provinces. The method allows the optimal detection of coherent 
oceanic sub-regions originating from the ocean circulation and its multiscale variability. 

The coherence ratios are generally elevated ($\rho \in [0.5,1]$) and variable (Fig. \ref{fig:snapshots}). 
Although it depends on both the local leaking processes and the area of a given province, there is no apparent 
relationship between the size of the sub-region and its coherence ratio. Overall, $\rho \ge 0.8$ are often seen in 
the Aegean and Adriatic Seas. The Alboran, Balearic, Tyrrhenian, and Adriatic Seas are characterized by relatively 
large provinces, whereas the Levantine, Aegean and south Ionian and Libyan Seas are subdivided in rather 
small ones. Note also that some provinces are composed of non-contiguous boxes. This occurs especially within 
the pathways of fast-flowing currents as the Algerian Current, the Atlantic-Ionian stream (south Ionian, Libyan and 
south-east Levantine sea) and the Liguro-Provencal Current (Ligurian sector).

\subsection{Spatial-scales and geographic structure of larval dispersal}
\label{climatoprovinces}

\begin{figure}
\noindent
\begin{center}
\includegraphics[width=0.5\textwidth, clip=true]{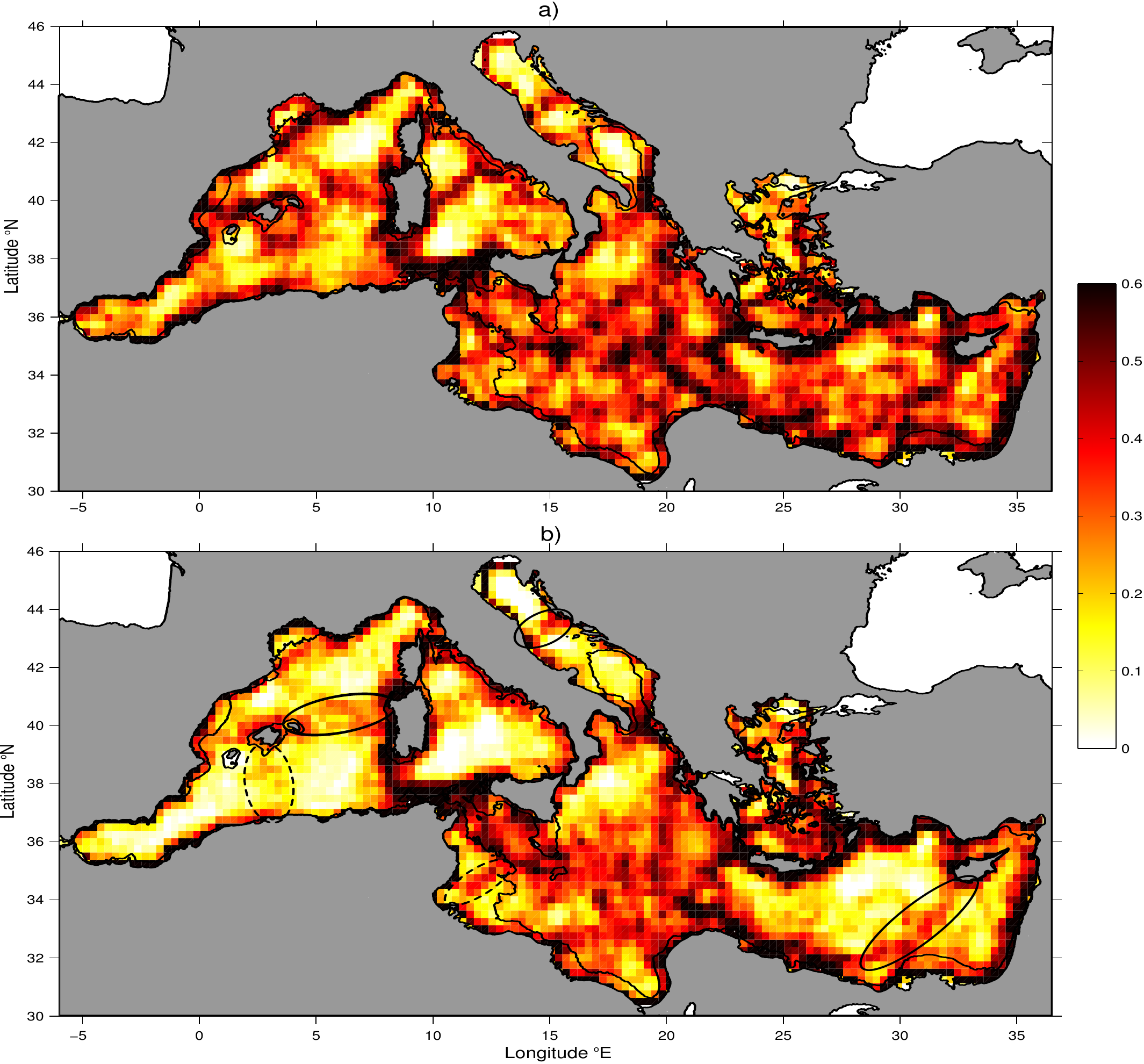}
\caption{Frequency of occurrence of province boundaries at each ocean node from the ensemble of 20 experiments (all winters/summers 
of $2002-2011$) for each PLD: a)  $\tau \simeq PLD = 30$ days and b) $\tau \simeq PLD = 60$ days. Black ellipses in b) highlight 
the frontiers which have significant seasonality: plain ellipses indicate a preferential occurrence in winter and dotted ellipses 
in summer. Black contours represent the 200 m isobath.}
\label{fig:provinces}
\end{center}
\end{figure}

The frequency of occurrence of province boundaries is now examined across the ensemble 
of experiments to identify recurrent frontal systems and relatively stable hydrodynamical units which would
organise larval dispersal. Over most coastal/shallow regions, boundaries occur in various locations 
and orientations, resulting in no apparent structure (dark red patches). These disorganized 
patterns characterize oceanic environments with complex circulation in which spatial-scales of connectivity 
are highly variable \cite{siegel2008}. They are observed in most insular regions (Balearic, Tuscan 
and Aegean archipelagos, Corsica, Sardinia, Crete, Cyprus), in the Tunisian-Sicilian 
strait (also punctuated by small islands) and over narrow continental shelves (Italian, French, 
Catalan, Libyan-Egyptian and Israelian-Lebanon shelves) (Fig. \ref{fig:provinces}a). 
In contrast, wide continental shelves are organized as coherent hydrodynamical units whose offshore 
limits match the 200 m isobath. The gulf of Lion is delimited by a frontier coinciding 
with the Catalan front and associated Northern current \cite{bouffard2010} (an extension of the Liguro-Provencal current)
. For long PLDs only, the Tunisian-Libyan shelf appears as two units in summer 
(Fig. \ref{fig:provinces}b), merging into a single one in winter. The oceanic frontiers 
constituted by such currents/fronts are likely to prevent coastal larvae from escaping wide 
continental shelves. 

In the open ocean, clear hydrodynamical units emerge (Fig. \ref{fig:provinces}), organized 
as large ``gyre'' systems with rare occurrence of boundaries (white/yellow colors) in their center 
and semi-persistent frontiers (light/dark red colors) aligned along their perimeters. Elevated 
connectivity within each subdivision but little exchange between them are expected, thus providing 
basin-scale patterns of larval dispersal. Large hydrodynamical units are 
found in the western Mediterranean basin, the Adriatic sea, the Tyrrhenian 
sea (Fig. \ref{fig:provinces}a and b) and only at longer time-scales in the 
north-Ionian and Levantine seas  (Fig. \ref{fig:provinces}b). Most of these 
open-ocean frontiers are located along well-known oceanographic features \cite{millot2005}, some of them 
recognized as partial transport barriers. For instance, the so-called Oran-Almeria front separates the 
Alboran sea from the rest of the Mediterranean Sea. It appears here rather extending from Oran to Cartagena, 
some 200 km away than previously documented \cite{mermex2011}. The Balearic 
front is another semi-permament transport barrier \cite{mancho2008} passing north of the Balearic 
archipelago in the Balearic current \cite{bouffard2010} and elongating eastward in winter. 
North of this quasi-zonal boundary, a large hydrodynamical unit composed of the 
Lion gyre and Ligurian sea is separated from the Balearic sea at short time-scale. The Tyrrhenian 
sea is consistently organized as a two-gyre system using both PLDs. For the 30-day integration the 
Adriatic sea is subdivided by bathymetric gradients ($\sim$ 100 and 200 m isobaths, i.e. off the Gargano 
promontory) into a northern, central and southern Adriatic gyres, the two latter units merging 
for PLDs of 60 days.  

Surprisingly, some open-ocean areas, such as the Ionian, Levantine and Aegean basins (Fig. \ref{fig:provinces}a), 
are characterized by disorganized dispersal patterns and stochastic larval connectivity \cite{siegel2008}. They 
become more structured at longer time-scales with the emergence of the Western Ionian gyre, the Shikmona gyre 
and a large system encompassing the Rhodes, Ierapetra and Mersa-Matruh gyres \cite{millot2005}. The eastern Aegean 
sea has disorganized dispersal patterns whereas small hydrodynamical units appear in its northern and western parts, 
in good agreement with its thermal structure \cite{poulos1997}. 

More generally, regions with no apparent spatial patterns at short PLDs see the emergence of 
spatial structures for longer integration time. Oceanic areas already identified as gyral systems 
for short time-scales have their diameter increasing with the integration time, ultimately merging with their neighbors. 

Note that most of these hydrodynamical units are quite consistent with the trophic clusters obtained from 
satellite chlorophyll data \cite{dortenzio2009}, suggesting they also delimits specific biogeochemical 
provinces \cite{longhurst2006}. Indeed, although this study focuses on passive larvae, 
the unveiling of well-known oceanic fronts and gyres hint that the spatial distribution of other tracers 
(e.g. salinity, temperature, chlorophyll-\textit{a}, dissolved nutrients) are also influenced by similar 
transport patterns.

\subsection{Implications for the design of marine reserves}
\label{mpaconnectivity}
 
\begin{figure}
\noindent
\begin{center}
\includegraphics[width=0.5\textwidth, clip=true]{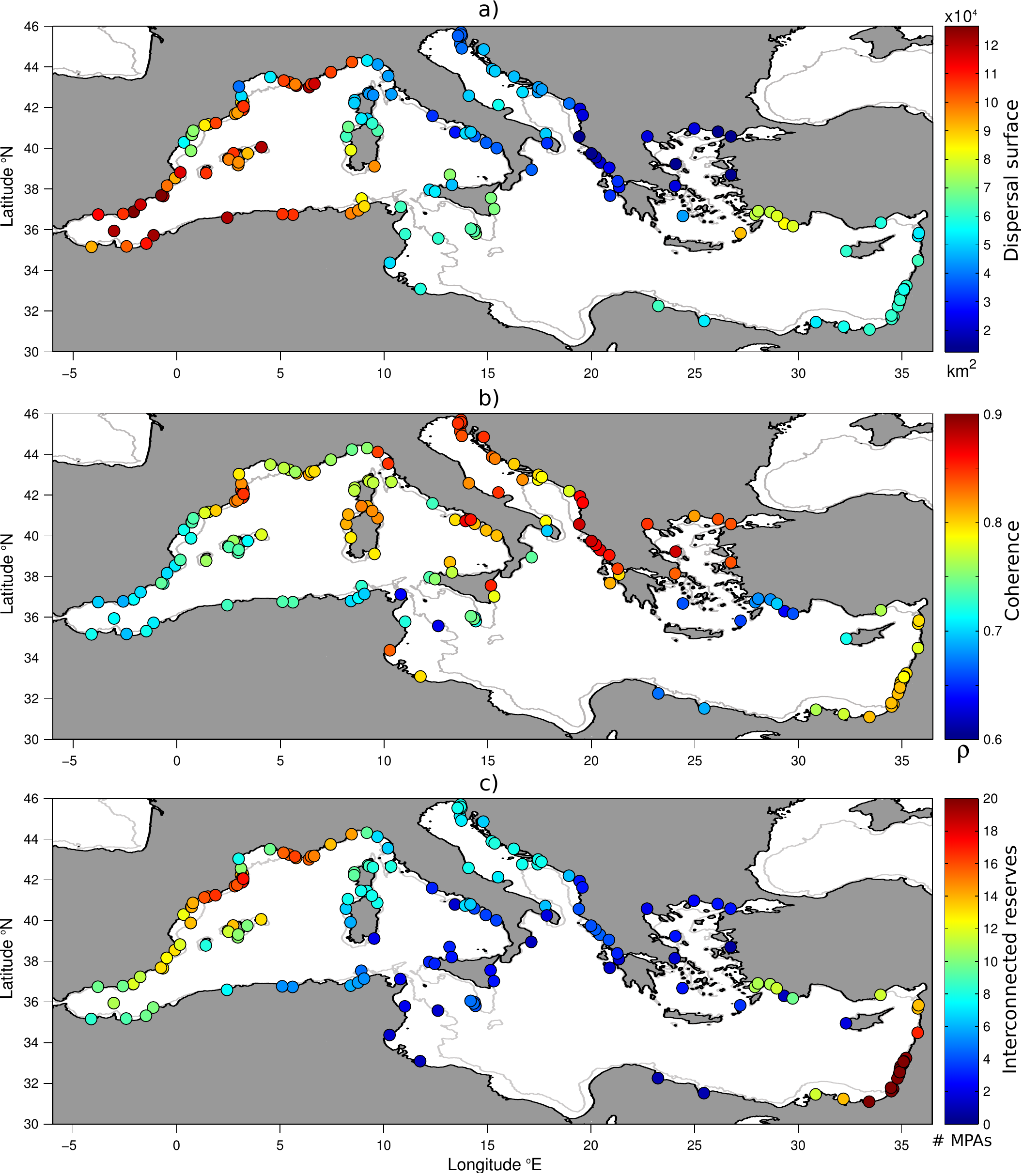}
\caption{Spatial variability of MPAs connectivity derived from three complementary metrics averaged over all 
winter/summer experiments over $2002-2011$ using $\tau \simeq PLD = 30$ days. a) Mean area (in $km^{2}$) 
and b) mean $\rho$ of the province sheltering the reserve of interest. c) Mean number of interconnected MPAs 
(i.e. number of reserves situated within the same hydrodynamical province). Light grey contours represent the 200 m 
isobath. Results using a PLD of $45$ and $60$ days are qualitatively similar with a slight increase of 
the mean area and the number of interconnected reserves. Note that non-contiguous areas belonging 
administratively to a given reserve were treated here as a single MPA.}
\label{fig:mpavar}
\end{center}
\end{figure}

The geographical structure of larval dispersal in the seascape influences largely the connectivity 
of marine reserves. The MPAs located within large and stable hydrodynamical units (Fig. \ref{fig:provinces}) 
are interconnected, in good agreement with \cite{andrello2013} who identified similar MPA clusters 
in the Algerian, Balearic, Adriatic and Tyrrhenian seas, respectively. Further information is 
obtained with the analysis of three complementary proxies of connectivity defined as followed. 
We analyze the mean spatial scales of larval dispersal (Fig. \ref{fig:mpavar}a) and the mean local 
coherence (inversely related to leaking, Fig. \ref{fig:mpavar}b) by averaging over the ensemble of 
experiments the area and the coherence $\rho$, respectively, of the time-dependent province encompassing 
each MPA. While these two metrics are solely influenced by the flow, the mean number of interconnected 
MPAs (i.e. temporally averaged number of MPAs encountered within the same time-dependent province, 
Fig. \ref{fig:mpavar}c) depends also on the density of existing reserves.

Larval connectivity and dispersal potentials are highly variable among the Mediterranean MPAs (Fig. \ref{fig:mpavar}). 
Reserves in the Adriatic and Aegean seas are characterized by small dispersal surface ($\le 5 \times 10^{4}$~$km^{2}$) 
and among the highest coherence ($\rho \ge 0.8$). This suggests a low connectivity which is also reflected in the 
few interconnected MPAs ($\le 8$) despite their relatively high density. MPAs located around isolated islands
are associated with modest dispersal surface ($\sim 4-8 \times 10^{4}$~$km^{2}$)
and low coherence ($\rho \le 0.7$). Typical of these insular environments \cite{vaz2013}, complex circulation patterns 
(islands' wake, eddies, retention...) result in a moderate connectivity and high temporal variability (not shown). 
MPAs implemented within narrow continental shelves bounded by energetic currents are characterized by rather 
large provinces ($\ge 7 \times 10^{4}$~$km^{2}$) and moderate coherence ($0.65 \le \rho \le 0.8$). These reserves 
are situated along the French C\^ote d'Azur with the Liguro-Provencal Current, the Catalan coast with 
the Northern Current, the Moroccan/Algerian coastlines impacted by the Algerian Current and in the eastern 
Levantine basin with the jet-like intensifications of its gyre circulation \cite{millot2005}. 
This elevated connectivity is driven by the adjacent currents that regularly intrude onto the shelf, 
enhancing larval dispersal along the current axis, as suggested by the numerous 
interconnected MPAs ($\ge 15$) along the French, Catalan and Israelian coastlines. In contrast, MPAs situated 
within extended continental shelves, such as the Gulf of Lion and the Tunisian/Libyan shelf, are 
characterized by small dispersal area ($\le 6 \times 10^{4}$~$km^{2}$) and large coherence ($\rho \ge 0.8$). 
Unless exceptional intrusion events (not shown), the inner-shelf is isolated by shallow bathymetry 
holding the current off the shelf break, thus resulting in restricted connectivity. Note that most MPAs 
associated with narrow shelves and sluggish circulation (such as the Tyrrhenian, Corsican and Sardinian 
coastlines) behave quite similarly to the latter group with small dispersal surfaces.

Despite the stochastic nature of larval dispersal \cite{siegel2008}, local oceanographic characteristics result 
in the emergence of connectivity regimes. They should be in accord with the main conservation objectives to ensure 
successful implementations of coastal and offshore marine reserves. For instance, the allocation of MPAs within 
narrow shelves bounded by currents would favor larval export over large distances \cite{pelc2010} whereas 
reserves created within internal seas or large continental shelves would rather promote the restoration of 
local populations \cite{pineda2007}. Overall, the Mediterranean MPAs are not evenly 
distributed across the spatial partitioning of the seascape revealed by our analysis (sect. 
\ref{climatoprovinces}) \cite{dejuan2012}. Moreover, the ``size and spacing'' 
guidelines already studied theoretically \cite{moffit2011}, may differ depending on the local dispersal 
behavior. Our results suggest the use of few large MPAs located in each stable hydrodynamical 
unit of the western Mediterranean basin and the Adriatic sea whereas numerous small MPAs 
evenly distributed across the fluctuating units might be preferable in the Ionian and Aegean seas.

\section{General conclusions and perspectives}

Using a method coupling Lagrangian trajectories and new tools from Network Theory, we study larval 
dispersal by surface currents in the Mediterranean Sea. Under our assumptions, a transport 
network is constructed from the horizontal advection of passive particles in a 
modelled oceanic flow, simulating larvae of different planktonic seasons and durations. The systematic detection 
of communities in the network extracts a set of hydrodynamical provinces which organize the surface
dispersion of larvae in the entire Mediterranean basin. Their boundaries coincide with both mesoscale 
and regional-scale oceanographic features, comprehending the multiscale processes of ocean circulation. 
The repeated occurrence of these frontiers allows separating the seascape in different hydrodynamical units which 
provide the ``backbone'' of oceanic transport impacting larval dispersal and connectivity among existing 
MPAs. While the role of such large-scale dispersal patterns on the genetic structure of marine population 
remain to be determined, the hydrodynamical units evidenced may be used to optimize the sampling 
strategy of genetic studies. The similarity between our flow-driven boundaries and major environmental 
gradients commonly used to regionalize the Mediterranean seascape finally suggests they might also define 
oceanic biomes or even faunistic units. Future developments of the methodology would
have to consider more realistic larval behavior for a given target species as well as full tridimensionality 
of the flow. These extensions may help incorporating large-scale biogeography and dispersal patterns to 
improve MPAs design toward efficient management and conservation of marine ecosystems.

\begin{acknowledgments}

The authors acknowledge support from MICINN and FEDER through the ESCOLA project (CTM2012-39025-C02-01) 
and support from ECs Marie-Curie ITN program (FP7-PEOPLE-2011-ITN) through the LINC project (no. 289447). 
The simulated velocity field used in this study was generated by MyOcean (\textit{http://www.myocean.eu/}) and the 
locations of Mediterranean MPAs were provided by MedPan (\textit{http://www.medpan.org/}). The authors thank the 
two anonymous reviewers who helped improving the original manuscript.

\end{acknowledgments}

\newpage
\section{SUPPLEMENTARY INFORMATION 1: Eddy-resolving model and discretization procedure}

\subsection{Hydrodynamical model of the Mediterranean Sea}

The Mediterranean Forecasting System is a hydrodynamic model based 
on NEMO-OPA (Nucleus for European Modelling of the Ocean-PArallelisé, 
version 3.2 \cite{madec2008}) with a variational data assimilation scheme. It 
is a primitive equations model in spherical coordinates, implemented in the 
Mediterranean at $\frac{1}{16}$ $deg$ horizontal resolution and 
72 unevenly spaced vertical levels \cite{oddo2009}. We use here the 
"Physics reanalysis" component for years $2002-2011$ downloaded from MyOcean website. 
 
The model covers entirely the Mediterranean Basin and extends into the 
Atlantic in order to better resolve the exchanges with the Atlantic 
Ocean at the Strait of Gibraltar. It is nested, in the Atlantic, within 
the monthly mean climatological fields computed from ten years of daily 
output of the  $\frac{1}{4}$ $deg$ global model \cite{drevillon2008}. 
Details on the nesting technique and major impacts on the model results 
can be found elsewhere \cite{oddo2009}. The model uses vertical partial cells
to fit the bottom depth shape. It is forced by momentum, water and heat 
fluxes interactively computed by bulk formula using the 6-h, 0.25 $deg$~horizontal-resolution 
operational analysis and forecast fields from the European Centre for 
Medium-Range Weather Forecasts. Air-sea processes predict surface temperature \cite{tonani2008}, while the 
water balance is computed as Evaporation minus Precipitation and Runoff. 
The evaporation is derived from the latent heat flux; the precipitation 
and the runoff are provided by monthly mean datasets. The Dardanelles 
inflow is parameterized as a river using climatological net inflow 
rates \cite{kourafalou2003}. 
 
The data assimilation system is the OCEANVAR scheme \cite{dobricic2008}. The background error correlation matrices, 
estimated from the temporal variability of parameters in a historical 
model simulation, vary seasonally in the sub-regions of the Mediterranean 
Sea characterized by different physical characteristics \cite{dobricic2007}.
The Mean Dynamic Topography is used for the assimilation of Sea Level Anomaly 
(SLA) \cite{dobricic2005}. The assimilated data include: 
along track SLA, satellite Sea Surface Temperature (SST), in-situ temperature 
profiles by eXpandable Bathy Thermograph, in-situ temperature and salinity 
profiles by Argo floats, in-situ temperature and salinity profiles from 
Conductivity-Temperature-Depth casts. Objective Analyses of SST data are 
used for the correction of surface heat fluxes with the relaxation constant 
of $40$ $Wm^{-2}K^{-1}$ 

\begin{figure}
\noindent
\begin{center}
\includegraphics[width=0.5\textwidth, clip=true]{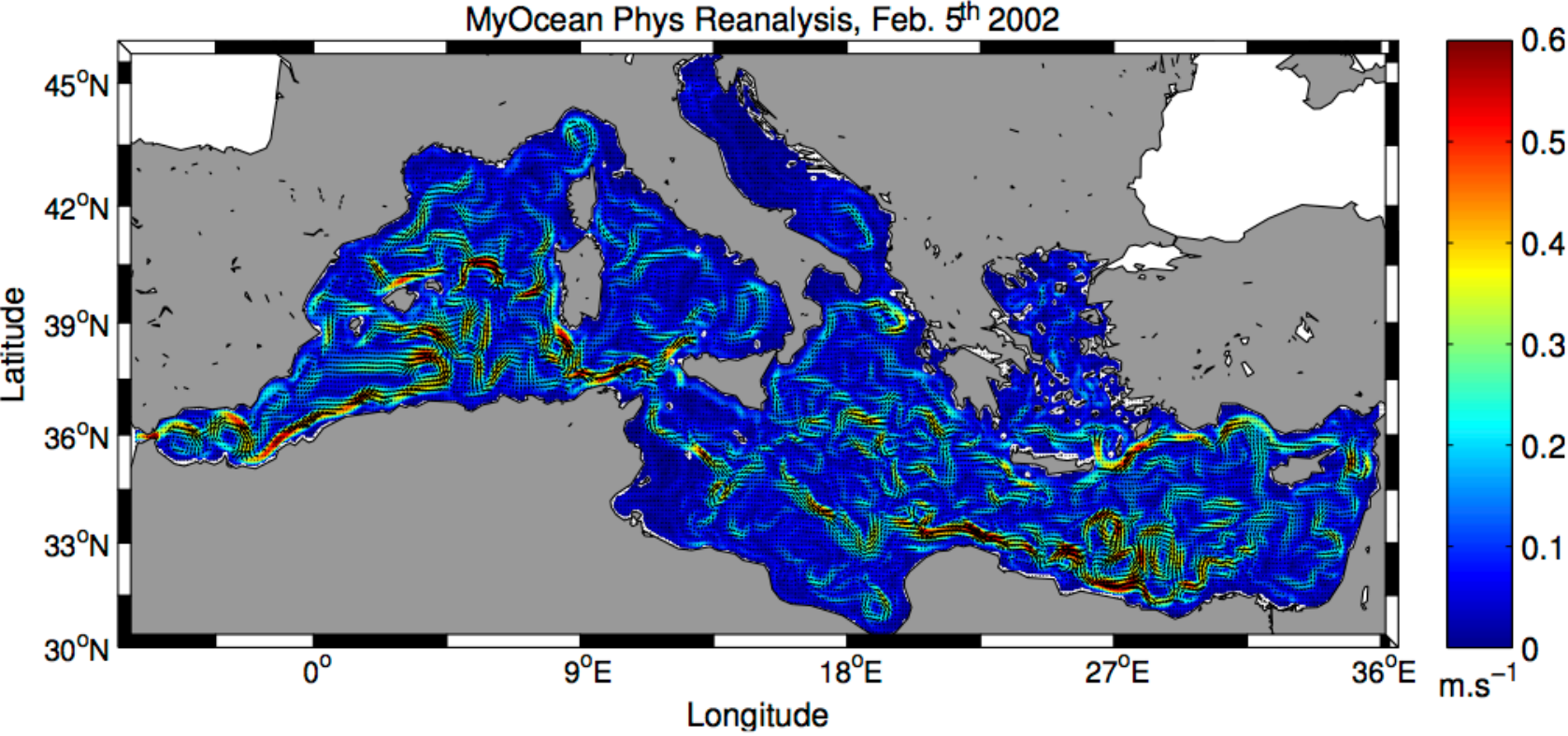}
\caption{Snapshot of the surface (8 m) velocity field for February $5^{th}$ $2002$. Background 
colors represent the modulus ($ms^{-1}$) of the instantaneous velocity field which is superimposed as 
black vectors (note that the original $^{1}/_{16}$ $deg$ resolution of the vector field has been coarse-grained 
for visualization purposes).}
\label{fig:snapshotNEMO}
\end{center}
\end{figure}

An exemplary snapshot of the surface (8 m) velocity field generated by
this configuration and used to integrate Lagrangian particle trajectories is 
displayed in Fig. SI-\ref{fig:snapshotNEMO}. Note the realistic representation 
of both large- and small-scale oceanographic features.

\subsection{Discretization procedure to define the network nodes}

\begin{figure}
\noindent
\begin{center}
\includegraphics[width=0.5\textwidth, clip=true]{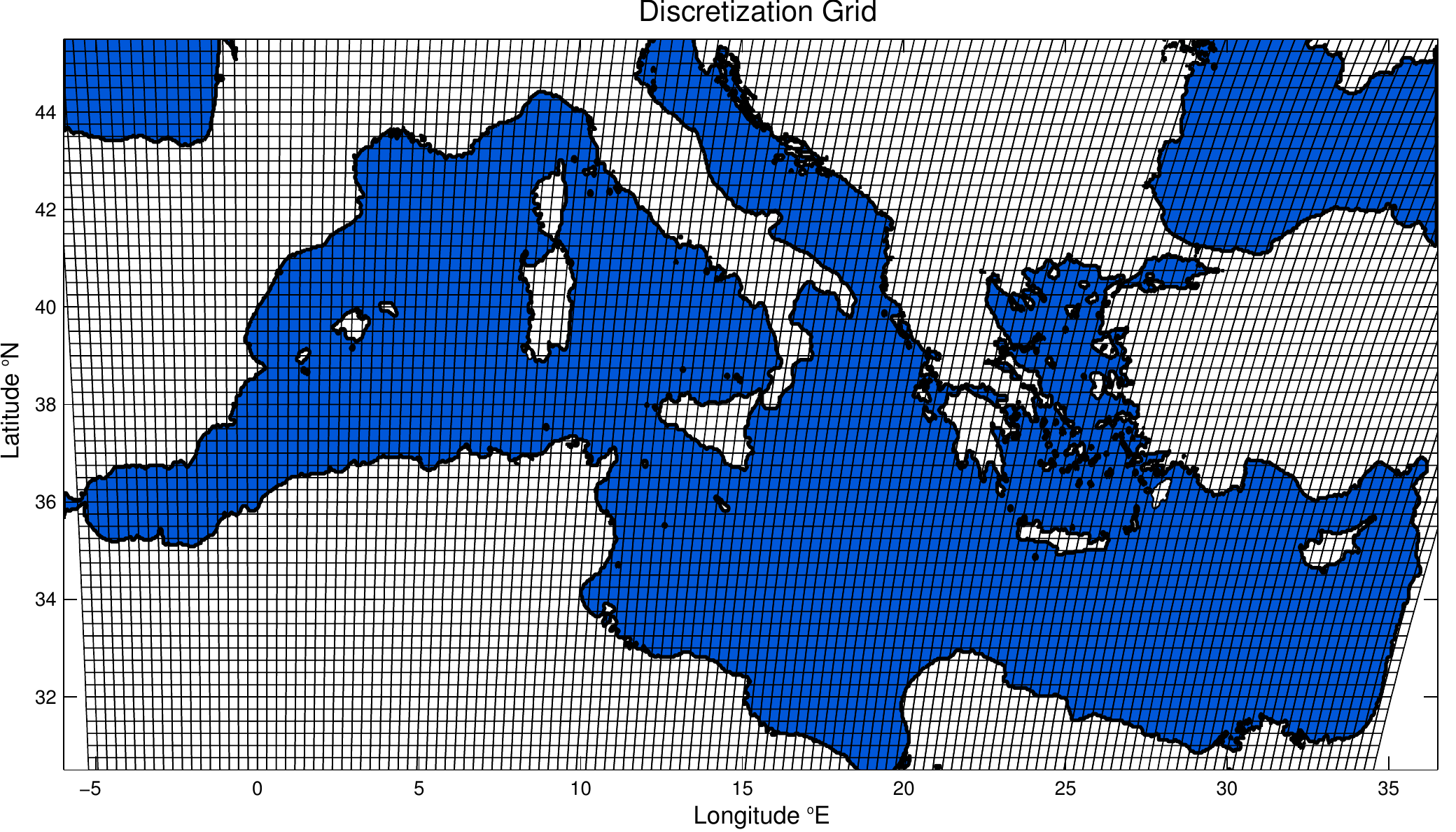}
\caption{Discretization grid used to construct the transport network. Each quasi-square box of $1/4$ $deg$ 
horizontal-resolution represents a node of the network.}
\label{fig:grid}
\end{center}
\end{figure}

The nodes of the transport network are delineated by discretizing the continuous ocean into quasi-square 
boxes of $1/4$ $deg$~horizontal-resolution, of the order of important mesoscale features of the 
Mediterranean circulation \cite{millot2005,mermex2011}. The ocean surface is subdivided into 
a total of $3270$ two-dimensional boxes, imposing equal-area ($\sim 772$~$km^{2}$) with the use of a 
sinusoidal projection (see Fig. SI-\ref{fig:grid}). This procedure introduces numerical diffusion below the scales 
of discretization \cite{froyland2003} so that sub-mesoscale dynamics are not explicitly resolved.

The simulated currents used to compute the Lagrangian trajectories (e.g. Fig. SI-\ref{fig:snapshotNEMO}) contain 
smaller scale structures since they were generated at $^{1}/_{16}$ $deg$, i.e. ~6-8 km at these latitudes. The box 
length-scale of $1/4$ $deg$~intervene in the community detection algorithm (so that the numerical diffusion 
introduced is impacting the spatial precision of their boundaries).

Adding noise (or random walk) to the Lagrangian trajectories to represent non-resolved small-scale processes 
is not necessary here, as our discretization procedure already introduces an implicit diffusion. It can be estimated
by considering that, away from strong fronts, the currents in the surface ocean are incompressible and two-dimensional, 
so that locally the flow is dominated by strain. Under the influence of this flow, patches of tracer (i.e. larvae,
chlorophyll-$a$, etc...) evolve towards filamental structures \cite{martin2000,abraham2000} of typical width $L=\sqrt{^{D}/_{\lambda}}$, 
where $D$ is the diffusivity coefficient and $\lambda$ the rate of strain (or Lyapunov exponent) of the flow. 
From this expression, one can estimate an effective diffusion coefficient $D=\lambda\,L^{2}$, associated to structures 
resolved by our discretization of size $L$. Since $\lambda$ spans $0.01 - 0.1$ $day^{-1}$ in the Mediterranean Sea 
\cite{dovidio2004} and  $L= 1/4$ $deg$~($\sim$ $25$ $km$ at these latitudes), this formula gives a numerical 
diffusivity O$(1 - 10^{2})$~$m^{2}s^{-1}$. 

Other estimations, e.g. based on Okubo's formula associated to a given resolution of the velocity field 
\cite{okubo1971,isma2011} provide values for the diffusivity of the same order of magnitude. Indeed 
eddy diffusivity coefficients in the Mediterranean Sea range from $1$ to $100$~$m^{2}s^{-1}$, while 
a value of $10$~$m^{2}.s^{-1}$ has been used in Lagrangian tracking model \cite{sayol2014}, falling precisely within the range of our estimations. 
This is also similar to the random noise typically added to Lagrangian trajectories. For instance a
diffusion of $0.2$~$m^{2}s^{-1}$ was used with a fine-scale numerical model at $\sim 4$ km spatial resolution \cite{vaz2013}, implying higher 
values (that would be within our estimated range) for a coarser model like the one we used. According to these estimates, 
the artificial diffusion introduced by our method is, in all cases, similar or even higher than the random noise typically 
added to Lagrangian trajectories to simulate unresolved sub-grid processes.

\newpage
\section{SUPPLEMENTARY INFORMATION 2: Ecological relevance of our numerical experiments}

The need for an ecosystem-based management of marine ressources has been emerging 
\cite{pikitch2004,kaplan2010b}, especially in the Mediterranean 
basin \cite{coll2012,coll2013,guidetti2013}. For instance, recent studies 
considered an ecosystem-based approach to optimize the location of reserves for several 
species based on multi-factorial analysis \cite{lagabrielle2012} or to inform MPA 
efficiency by modelling trophic interactions within the whole ecosystem \cite{colleter2012}. 
In the context of assessing larval connectivity for MPAs design, it implies the need of
considering the ecosystem as a whole rather than focusing on a specific organism 
\cite{guizien2006}. This is implemented here by simulating a range of 
different PLDs and periods of spawning which are, under certain assumptions, 
biologically relevant for a number of Mediterranean species. 
 
A compilation of the mean Pelagic Larval Durations (PLDs) of 62 littoral
Mediterranean fishes revealed they span 10-70 days depending on the species 
considered, with large intra-species variability \cite{macpherson2006}. Considering the "basin-scale" 
angle of our study, we focus on species with a wide geographical range and 
potential for large-distance dispersal.  
These organisms are usually characterized by pelagic spawning, long PLDs
($\geq$ 20 days) and offshore larval distribution (although many combinations of
such early life traits exist) \cite{macpherson2006}. We retain a PLD of 30 days which is the 
best estimate available for a few iconic species of the Mediterranean ecosystem,
including some demersal fishes (e.g. the groper \textit{Epinephelus marginatus} 
\cite{andrello2013}, the blenny \textit{Lipophrys canevai}, the wrasse \textit{Labrus viridis}, 
the goatfish \textit{Mullus surmuletus}, the bream \textit{Sarpa salpa} \cite{macpherson2006}) 
and invertebrates (e.g. the crab \textit{Pachygrapsus marmoratus} \cite{fratini2013}). 
We also performed simulations for a PLD of 60 days since other Mediterranean 
fishes (e.g. the blenny \textit{Lipophrys trigloides} \cite{macpherson2006})
and most marine invertebrates (echinoderms like the sea-star \textit{Astropecten 
aranciacus} \cite{zulliger2009}, some molluscs and many exploited crustaceans 
\cite{shanks2009}) are characterized by long PLD. 
     
Dispersal potential can also be influenced by other mechanisms than PLD, such 
as early life traits. However, because the precise description of spawning strategy 
and larval distribution of marine organisms remains elusive,
a classification was proposed based on its preferential season and location of 
occurrence  \cite{macpherson2006}. A large majority of the species they studied (e.g. \textit{Epinephelus marginatus}, 
\textit{Lipophrys canevai}, \textit{Mullus surmuletus}) spawn in late spring / early summer, 
so we assume that summer is the season of their planktonic life. Some others (e.g. \textit{Sarpa salpa}, 
\textit{Lipophrys trigloides}) prefer a late autumn / early winter spawning, resulting 
in the highest abundance of larvae observed in winter. Concerning spatial preferences, the "coastal 
spawners" (e.g. \textit{Labrus viridis}) release their eggs close to the bottom in shallow areas
and then their planktonic larvae are concentrated in the coastal ocean. In contrast, the "pelagic 
spawners" (e.g. \textit{Epinephelus marginatus}, \textit{Mullus surmuletus}, \textit{Sarpa salpa}) 
spawn in the open-ocean with their larvae found widespread offshore \cite{macpherson2006}.
  
To sum-up, an ensemble of Lagrangian simulations covering both coastal and open-ocean regions 
were carried-out for each PLD over all years $2002-2011$ with two starting dates, a winter 
($t_{0}$ = $1^{st}$ Jan.) and a summer ($t_{0}$ = $1^{st}$ Jul.) calculations. Our basin-scale 
patterns of connectivity (Fig. 2 and 3) are extracted from the temporally-averaged hydrodynamical 
provinces (thus erasing the impact of the inter-annual variability of the circulation), i.e. 
derived from more than 30 millions trajectories per PLD. It leads to statistically robust 
and ecologically meaningful patterns of larval connectivity in the Mediterranean 
basin, providing useful results in the context of an ecosystem-based MPA design.

\newpage
\section{SUPPLEMENTARY INFORMATION 3:  The \textit{Infomap} community detection method}
\subsection{Why using \textit{Infomap}?}

Once the transport problem is cast in terms of network concepts by
constructing the connectivity matrix $\textbf{P}^{t_0,\tau}$, a
variety of mathematical tools becomes available for the question 
of dividing the sea into well-connected regions called ``communities'' in the 
network theory parlance \cite{newman2010,danon2005,lancichinetti2009,fortunato2010}. 
Among the numerous methods available, we choose \textit{Infomap} \cite{rosvall2008}, 
a well known algorithm in the Network community, to 
identify community structure in our hydrodynamic networks. It retains both the 
``direction'' and ``weight'' information of each link by mapping the system-wide flow 
induced by local interactions between nodes. This is a key feature of the method 
since the weights (amount of water transported) and directions (of the flow) in our 
transport network are crucial to represent the true oceanic circulation. 

This flow-based approach is indeed necessary to identify the most important 
structural aspects of networks where links represent patterns of movement among 
nodes (such as the transport network built here). In contrast, other topological 
methods which also use ``direction'' and ``weight'' (i.e. modularity 
optimization or cluster-based compression) are well adapted to analyse
networks whose links do not represent flows but rather pairwise relationships, 
since they are blind to interdependence in flow networks \cite{rosvall2008}. 
In addition, \textit{Infomap} does not require to
specify {\sl a priory} the number of communities to be
detected. It finds structures which are directly related to
well-mixed regions under the flow represented by
$\textbf{P}^{t_0,\tau}$, and not to other structural properties
(for example, a well defined region with strong fluxes oriented
towards a particular direction) which will not lead to particle
localization in that region. Also, \textit{Infomap} does not assume
communities with similar sizes (as for example the spectral
partitioning \cite{froyland2007}) nor suffers from the
``resolution limit'' \cite{fortunato2007} which limits the minimum
community size detectable by most algorithms. In fact, the transport network 
is decomposed into pieces of different sizes in regions where the 
flow requires so (e.g. Fig. 1). 

All these properties make it very suitable for our purpose of identifying 
well mixed oceanic regions which are relatively less connected 
with the surroundings. Last but not least, the minimization algorithm is computationally efficient, 
well documented and publicly available (http://www.tp.umu.se/\verb+~+rosvall/code.html).

\subsection{How it works?}

\textit{Infomap} considers an ensemble of random walkers in the network
caracterized by $\textbf{P}^{t_0,\tau}$, moving with the
transition probabilities in that matrix. Then, the method
addresses from the information-theory point of view the
question of optimally coding the ensemble of possible random
walks. To this end the network is divided in communities and
each random walk is coded by sequences of words that represent
successive locations inside a community and jumps to a
different community. The information-theoretic lower bound to
the average length of the codeword used is given in terms of
the transition probabilities and of the specific partition in
communities by the so-called {\sl map equation}. For a network
characterized by a directed and weighted adjacency matrix, equivalent to 
our connectivity matrix $\textbf{P}^{t_0,\tau}$, this map equation is:

\begin{equation}
L=q_\curvearrowright  H({\cal Q}) + \sum_{\alpha=1}^c p_\circlearrowright^\alpha H({\cal P^\alpha})\ .
\label{map}
\end{equation}

$c$ is the number of communities in the particular partition
considered. The first term involves the Shanon entropy
associated to the transitions between different communities
$\alpha$:

\begin{equation}
H({\cal Q})= -\sum_{\alpha=1}^c \frac{q_{\alpha\curvearrowright}}{q_\curvearrowright}
\log_2\left(\frac{q_{\alpha\curvearrowright}}{q_\curvearrowright}\right)
\end{equation}

$q_{\alpha\curvearrowright}$ is the probability to leave
community $\alpha$ in one random-walk step, and
$q_\curvearrowright=\sum_{\alpha=1}^c
q_{\alpha\curvearrowright}$. Expressions for these quantities
in terms of the components of the network matrix
$\textbf{P}^{t_0,\tau}$ exist \cite{rosvall2008}. The second term
in Eq. (\ref{map}) contains the Shanon entropies $H({\cal
P^\alpha})$ associated to the words used to codify the position
inside a community $\alpha$ and the word that denote the exit
from that community:

\begin{equation}
H({\cal P^\alpha})= -\sum_{i\in\alpha} \frac{\pi_i}{p_\circlearrowright^\alpha}
\log_2\left(\frac{\pi_i}{p_\circlearrowright^\alpha}\right)- \frac{q_{\alpha\curvearrowright}}{p_\circlearrowright^\alpha}
\log_2\left(\frac{q_{\alpha\curvearrowright}}{p_\circlearrowright^\alpha}\right)\ .
\end{equation}

The notation $i\in\alpha$ indicates sum over the nodes
pertaining to community $\alpha$. $\pi_i$ is the stationary
distribution of the random walk and
$p_\circlearrowright^\alpha=q_{\alpha\curvearrowright}+\sum_{i\in\alpha}
\pi_i$. Again, expressions for these quantities can be obtained from the 
elements in the network matrix $\textbf{P}^{t_0,\tau}$ \cite{rosvall2008}.

\textit{Infomap} finds the partition that minimizes the quantity in
(\ref{map}), i.e. the partition that provides a shorter
description of the ensemble of walks going in and outside the
communities. In other words, it finds the partition for which the random walks 
remain most of the time inside the communities with few jumps between them. 
This minimization process uses a deterministic Greedy algorithm followed by 
a simulated-annealing algorithm which was repeated $100$ times to select the 
best partition in provinces (although the results were already stable after 
$10$ attempts).


\end{document}